# Sensitive label-free and compact ultrasonic sensor based on double silicon-on-insulator slot micro-ring resonators

**Cheng Mei Zhang[a], Chao Ying Zhao[b,c]***

[a]*Nokia Solutions and Networks, Hangzhou, 310053, China*

[b] *College of Science, Hangzhou Dianzi University, Zhejiang 310018, China*

[c]*State Key Laboratory of Quantum Optics and Quantum Optics Devices, Shanxi University, Taiyuan 030006, China*

**ABSTRACT:** We propose a new label-free ultrasonic sensor, which comprises a slot wave-guide and double silicon-on-insulator (SOI) slot micro-ring resonators. The all-optical sensors do not suffer from electromagnetic interference. We choose to integrate a silicon slot double micro-ring (SDMR) resonators in an acoustically resonant membrane. Optimization of the several key structural parameters is investigated to achieve the mode-field distributions of transmission spectrum based on Comsol Multiphysics software. Our numerical studies show that the proposed ultrasonic sensor offers higher sensitivity and a larger detection frequency range than conventional piezoelectric based ultrasound transducer. For a SDMR system with an area of $15\mu m \times 30 \mu m$, sensitivity as high as $2453.7 \, mV/kPa$, and over a bandwidth range of $1\text{-}150 MHz$. The sensitivity value is 36 times higher than that of single slot micro-ring ultrasonic sensor. The theoretical $Q$-factor of the SDMR can be approximately $1.24 \times 10^6$ with bending radius of $5\mu m$. The investigation on the SDMR system is a valuable exploration of the photo-acoustic microscopy for the ultra-high $Q$ factor and large frequency range.

---

*Corresponding author.

E-mail address: zchy49@hdu.edu.cn.

# 1. Introduction

Integrated silicon photonic devices are widely applied in optical sensing, including bio-medicine and environment monitoring. Silicon photonic devices can offer high compatibility with the complementary metal-oxide-semiconductor (CMOS) fabrication processes and silicon-on-insulator (SOI) technology[1]. The various optical structures have been widely investigated, including Bragg gratings[2], micro-ring resonators [3], photonic crystal micro-cavities[4-5].

The low-cost, high-sensitivity, and label-free miniaturized ultrasound sensor is a key factor to obtaining high-resolution medical images and reducing misdiagnosis during the treatment of diseases. Fabrication of miniaturized ultrasound sensor that have a size below half of the wavelength of the ultrasound is challenging with conventional piezoelectric based ultrasound transducer, which suffers from geometry-dependent electrical noise and results in both poor depth resolution and inaccurate quantification of optical absorption information. In recent years, new ultrasound detection methods were put forward, there are four categories: (1) free-space-optics-based, including etalons[6], dielectric multilayer interference filters[7], Mach-Zehnder interferometer[8];(2) fiber-optics-based, such as fiber gratings[9]; (3) photonic integrated circuits, including micro-ring resonators[10], micro-sphere resonators[11], micro-disk resonators[12], and optics-fluids micro-ring resonators[13]; and (4) optical-interface-based, including surface plasmon resonance (SPR)[14], and metamaterials[15]. Considering the polymer micro-ring doesn't suffer from electromagnetic interference, the absorption of polymer is very small when the incident wavelength around $780\text{nm}$. Over the last decade, Guo and co-workers have reported on ultrasound sensor that are based on a polymer micro-ring resonator [16]. In 2012, Xie et.al proposed a possible to use the photo-acoustic pressure perturbation to generate ultrasound in the tissue to be photo-acoustic imaging. A micro-ring resonator with radius of $30\mu m$ and a cross section of $1.4\mu m \times 1.4\mu m$, the refractive index in the coupling medium is modulated, an ultra-high $Q$-factor of $3\times 10^5$ and a low noise-equivalent pressure (NEP) of 20Pa over 75MHz bandwidth were obtained [17].

A important class of ultrasound sensors are based on slotted wave-guides having an increased sensitivity. The concept of slot wave-guide was first proposed in 2004[18]. Between two high-index wave-guide layers, there is a wave-guide slot of tens of nanometeres with material of low index of refraction. The electric field discontinuity at the interface between high index contrast materials enables high optical confinement inside a gap region of low index material. Using slot wave-guide structures in the micro-ring resonators can improve the sensitivity, has a better $Q$ -factor[19-20]. They have a lot of applications such as refractive index sensing [21], optical manipulation of bio-molecules [22], and all-optical signal processing [23]. In 2017, a novel slot SU8 polymer micro-ring resonator ultrasound detector was proposed [24], the sensitivity is 165.7 times larger than conventional piezoelectric based ultrasound transducer. Considering the micro-ring ultrasonic sensor is usually in an aqueous environment, air-coupled non-contact ultrasound detection using capillary-based optical ring resonators was first investigated [25].

In this study, considering one micro-ring resonator has inherent limitations in the pass band structure, series coupling or parallel coupling structure double micro-ring resonator was put forward[26]. We propose the use of slot double micro-ring (SDMR) resonators to sense ultrasound. The basic idea here is based on the same principle as that of Ref. [27] and Ref. [28] and demonstrate that the structure employed in Ref.[29]. For slot micro-ring resonator, the optical power is trapped into slot wave-guides, which leads to an increase in the effective refractive index of the medium surrounding SDMR system. The SDMR system can be integrated in a mechanical structure to enhance sound-induced deformation and hence shift its optical resonance wavelengths. The ultrasound pressure will affect the optical mode of the slot wave-guide, the influence accumulates during the circulation of the light wave along SDMR system. A 3D FDTD simulations are performed and the sensing performances observed for optimize geometrical design parameters. We evaluate the performance of our sensor with varying sound pressures and light powers.

## 2. Principle and sensor structure

Fig.1 shows the schematic diagram of the proposed SDMR sensor designed on SOI technology. A 3D model of the input wave-guide and slotted ring coupling was constructed in COMSOL. In order to obtain outstanding performance of the SDMR system, the geometry of the slot wave-guide and the SDMR system should be optimized. The dimensional parameters of system (slot width $W_{slot} = 75\text{nm}$, slot height $H_{slot} = 280\text{nm}$, bus wave-guide width $W_{straight} = 350\text{nm}$, bus wave-guide height $H_{straight} = 280\text{nm}$, bus wave-guide/micro-ring1 gap $g_1 = 190nm$, micro-ring1/micro-ring2 gap $g_2 = 290\text{nm}$). Each slotted micro-ring1/micro-ring2 has a mean radius of $R = 5\mu m$ to the centre of the slot. The refractive index of silica $n_{sio_2} = 1.44$. The novel dual resonator design in this paper is configured such that any structural induced variations will affect the two resonators concurrently.

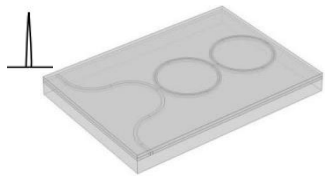
(a)
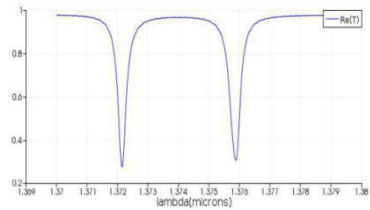
(b)

**Fig.1.** (a) Showing the micro-ring resonator on top of a membrane. The silicon dioxide top cladding to isolate the wave-guide from the water. Light transmitted through wave-guide is coupled to the resonator. (b) The optical transmission spectrum of SDMR ultrasonic sensor.

The slot wave-guide is constructed with two bus wave-guide with the gap embedded by GaN, which is serving as light input and output ports. The refractive index of the top layer $n_{GaN} = 2.3$. A quasi-TE mode incident beam is injected into the SDMR system through the input port as presented in Fig.1(a). As the transmission spectrum of the two ring resonators is the product of the transmission spectra of the individual resonators, it will exhibit two resonance peaks of the respective ring resonators. The $Q$-factor can be as high as $1.24 \times 10^6$ in Fig.1(b).

Based on optical resonance, the change in wave-guide cross section directly alters $n_{eff}$ of the guided mode, which results in a shift in the resonant frequency that can be conveniently monitored by measuring the modulation of transmitted optical signal through the bus wave-guide.[30]

$$T = \frac{t^2 + \alpha^2 \tau^2 - 2t\alpha |\tau| \cos[\phi + \phi_{neff}]}{1 + t^2 \alpha^2 \tau^2 - 2t\alpha |\tau| \cos[\phi + \phi_{neff}]} \quad (1)$$

Where $t$ is the transmission coefficient between bus wave-guide and micro-ring resonator, $\phi$ is the phase shift after light circulating along micro-ring, $\alpha$ is the amplitude transmission coefficient, and $\tau = (t - \alpha e^{i\phi})/(1 - t\alpha e^{i\phi})$ is the complex transmittance after through micro-ring 1. $\phi_{neff}$ is the effective phase shift of double micro-rings.

### 3. Power flow distribution and TE-mode distribution

The slotted ring design was adopted for this model in order to reduce the bend loss in the rings and so confine the circulating fields to central slots of rings.

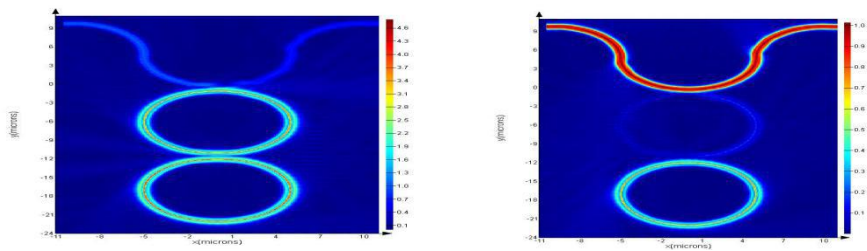

(a)             (b)

**Fig.2.** Power flow distribution of the quasi-TE mode at resonant wavelength (a) $\lambda = 1372.2\text{nm}$, (b) $\lambda = 1373.8\text{nm}$.

The coupling loss matches with the intrinsic loss in the micro-ring when resonance occurs, the propagation field energy is well confined in SDMR system when wavelength around 1372.2nm and 1373.8nm is shown in Figs.2(a)-(b), respectively. The light wave confined within the slot wave-guides, the total internal reflection of the light propagating in the SDMR system interacts with cladding via an evanescent tail.

We investigate numerically the quasi-TE fundamental mode confined in the SDMR structure based on 3D FEM method. We take the radius of micro-ring resonator, width of slot, width of wave-guide and the gap are set to $5\mu m$, 75nm, 350nm and 290nm, respectively. The micro-ring and the bus wave-guide have the same width, implying these are single mode around the operational wavelength of 1550nm.

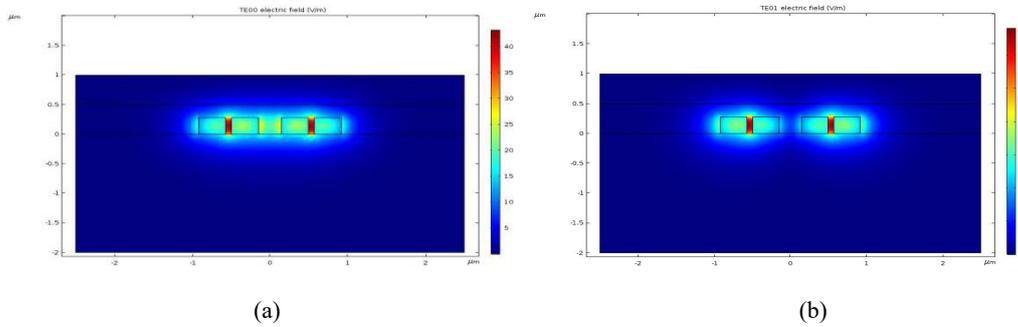

(a)             (b)

**Fig.3.** (a)TE00 mode distribution with a width of 350nm, (b)TE01 mode distribution with a width of 350nm.

As shown in Figs.3 (a)-(b), the electric field in the immediate vicinity of the silicon rails is greatly enhanced. This can be explained by large discontinuity of the $E$-field at the interface. The SDMR system can only support fundamental TE0 mode.

### 4. The principle and characteristic of ultrasonic sensor

The incident acoustic wave induces a strain field, the micro-ring resonator is sensitive to deformation which may be expressed in terms of strain (relative elongation)[31]

$$\varepsilon = -\frac{(1+\upsilon)(1-2\upsilon)P}{E}$$

(2)

where $\upsilon$ is Poisson's ratio, $E$ is Young's modulus, and $P$ is ultrasound pressure. The GaN material embedded in Silicon slot wave-guide. The Silicon slot wave-guide is above $SiO_2$ membrane and $SiO_2$ BOX layer. The acoustic resonance frequency of the $SiO_2$ membrane is determined by the material properties. Comparison of material properties of $SiO_2$ and GaN in Table 1.

| Material | density | Young's modulus | Poisson's ratio |
|---|---|---|---|
| $SiO_2$ | 2.65g/cm³ | 74.8GPa | 0.19 |
| GaN | 6.15g/cm³ | 181GPa | 0.352 |

**Table 1.** Comparison of material properties of $SiO_2$ and GaN.

In general, the performance of an ultrasound detector is characterized by the detection limit, $Q$-factor ($Q = \partial I/\partial P$) and sensitivity ($S = \partial T/\partial P$).

We assume that the SDMR structure is completely submerged in deionized water with refractive index of 1.333. The water layer serving as the medium for ultrasound wave incident from the top of GaN cladding. The incident ultrasonic wave slightly deforming the SDMR resonator, we choose to use the $SiO_2$ membrane to achieve a large deformation of the SDMR resonator.

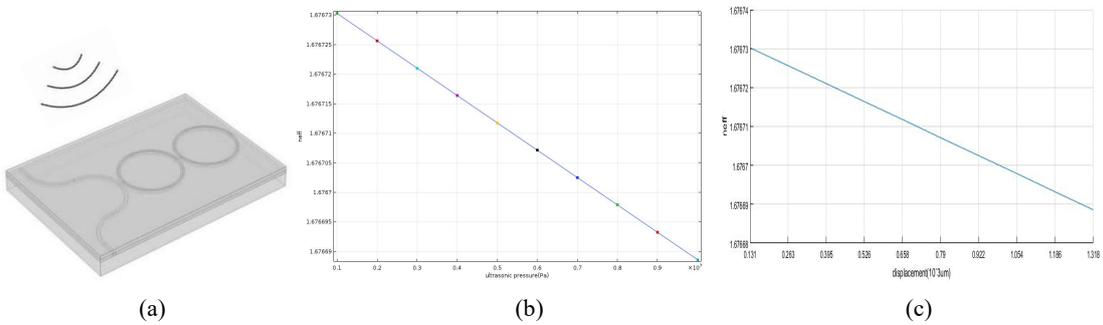

(a)　　　　　　　　　　　(b)　　　　　　　　　　　(c)

**Fig. 4.** (a) Sketch of the new type of ultrasonic sensor. (b) The $n_{eff}$ as a function of ultrasound pressure.　(c) The $n_{eff}$ as a function of displacement.

As shown in Fig. 4(a), the SDMR system for ultrasonic sensing. The influence of ultrasound pressure changes from 1MPa to 10MPa on $n_{eff}$ is illustrated in Fig.4(b). The influence of displacement on $n_{eff}$ is illustrated in Fig.4(c).

The deformation causes a change in circumference, a change in cross-section of wave-guide and a change in $n_{eff}$ of wave-guide and cladding (due to the photo-elastic effect), these changes result in a shift of optical resonance [20]. The shift of resonance wavelength and hence a direct measure of the deflection off the membrane. The relation between dimensions of the membrane and the pressure-induced strain in the SDMR resonator is nontrivial.

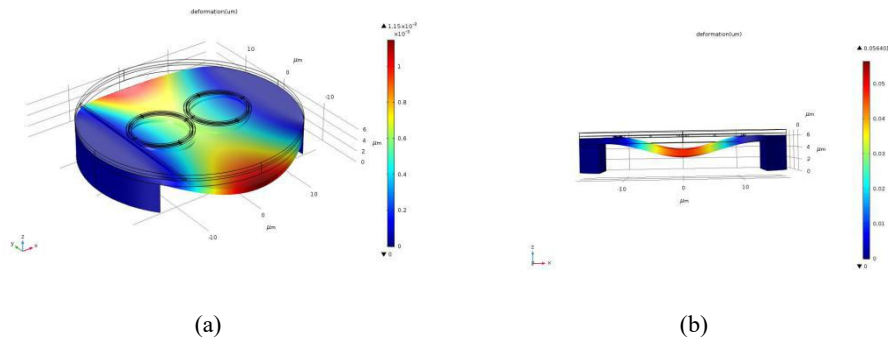

(a)                        (b)

**Fig.5.** (a) Sketch of the pressure-induced deformation and the strain for membrane with different ultrasound pressure. (b) The pressure above the membrane is 1MPa. Strain and deflection are computed using FEM simulation for COMSOL. The maximum deflection of the membrane is $0.056\mu m$.

We numerically simulated the pressure-induced deformation of the $SiO_2$ acoustical membrane using a FEM (implemented in COMSOL Multiphysics, see Fig.5(a)). The curvature of the $SiO_2$ membrane results in a position-dependent strain that scales linearly with the distance (see Fig. 5(b)). We record the strain at a point $0.056\mu m$ below the surface of the $SiO_2$ membrane at the center of the $SiO_2$ membrane.

## 5. The principle and characteristic of acoustic sensor

It is shown that a combination of optical and acoustic field measurements was performed using COMSOL Multiphysics software. Light is sent via optical fiber to sensor, then incident ultrasonic wave slightly deforming the SDMR resonator.

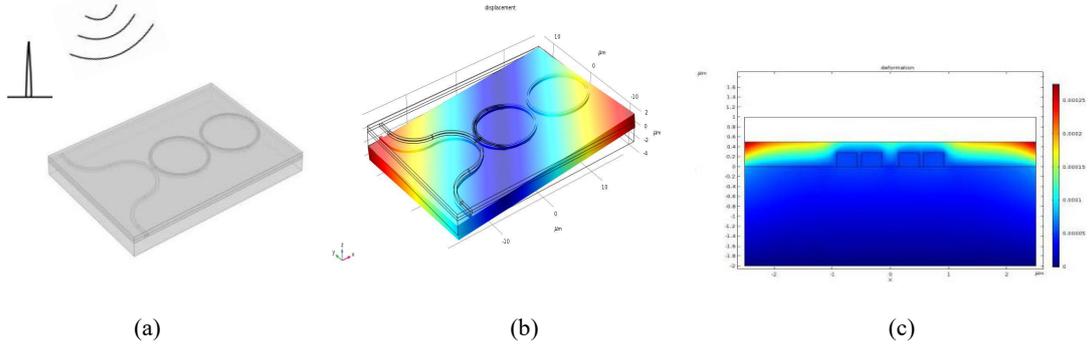

(a)          (b)          (c)

**Fig.6.** (a) Light is sent via optical fiber to sensor, then incident ultrasonic wave slightly deformed the SDMR system. (b)The deformation of the cross-section of the slot wave-guide region under 1MPa ultrasound wave.(c)The enlarged view of the deformation.

The slightly deformed the SDMR system due to optical-acoustic pressure perturbation, as shown in Figs. 6(a)-(c). The performance characteristics of some integrated SOI based ultrasonic sensors are compared in Table 2.

Some features comparison of SOI based sensor.

| Structure | $S(mV/kPa)$ | $NEP(Pa)$ | Q-factor | $R(\mu m)$ | $Feq(MHz)$ |
|---|---|---|---|---|---|
| MZI ring [8] | _ | 1.2 | $1.2706 \times 10^4$ | 5 | 0.77 |
| Polymer ring[32] | 36.3 | 88 | $10^7$ | 30 | 75 |
| SU8 ring [24] | 66.7 | _ | $8.34 \times 10^8$ | 12 | 540 |
| This work | 2453.7 | _ | $1.24 \times 10^6$ | 5 | 150 |

**Table 2**. Some features comparison of slot-based micro-ring sensors.

## 4. Conclusion

In this paper, we demonstrated that slot wave-guide coupled to SDMR system can be employed to sense strain or ultrasound. We demonstrated a new type of ultrasound sensor that has 36 times larger than that of single slot micro-ring ultrasound transducer. With an optimal design by COMSOL Multiphysics software simulations, detection sensitivity of $2453.7 mV/kPa$ and $Q$-factor of $1.24 \times 10^6$ can be achieved. Besides, our sensor with a compact footprint of $15 \mu m \times 30 \mu m$ and broad frequency range $150 MHz$ compared to traditional resonator. We believe that our novel ultrasound detector is a breakthrough for ultrasound array technology.

**Acknowledgments**

This work was supported by National Natural Science Foundation of China (grant number 11504074) and the State Key Laboratory of Quantum Optics and Quantum Optics Devices, Shan xi University, Shan xi, China (grant number KF201801).